\documentstyle[emulateapj,epsf]{article}
\submitted{slugcomment}

\def\sun{\hbox{$\odot$}}

\def\lesssim{\mathrel{\hbox{\rlap{\hbox{\lower4pt\hbox{$\sim$}}}\hbox{$<$}}}}
\def\gtrsim{\mathrel{\hbox{\rlap{\hbox{\lower4pt\hbox{$\sim$}}}\hbox{$>$}}}}

\def\arcdeg{\hbox{$^\circ$}}

\let\la=\lesssim
\let\ga=\gtrsim
\newcommand{\mm}[1]{\mbox{$#1$}}
\newcommand{\unit}[1]{\ifmmode \:\mbox{\rm #1}\else \mbox{#1}\fi}

\newcommand{\expec}[1]{\mm{\left\langle #1 \right\rangle}}
\newcommand{\mone}{\mm{^{-1}}}

\newcommand{\kms}{\unit{km~s\mone}}

\newcommand{\mpc}{\unit{Mpc}}

\newcommand{\hmpc}{\mm{h\mone}\mpc}

\newcommand{\lb}[2]{\mm{l = #1\arcdeg}, \mm{b = #2\arcdeg}}

\newcommand{\eqref}[1]{equation~(\ref{eq:#1})}

\newcommand{\figref}[1]{Fig.~\ref{fig:#1}}

\newcommand{\old}[1]{}
\newcommand{\sig}{\mbox{$\sigma$}}
\newcommand{\mg}{\mbox{Mg$_2$}}

\newcommand{\mgsig}{{\mg-- \sig}}

\slugcomment{ApJ Letters, in press}

\lefthead{M. J. HUDSON et al.}
\righthead{Large-Scale Bulk Flow of Clusters}

\begin{document}
\title{A Large-Scale Bulk Flow of Galaxy Clusters}

\author{
Michael J. Hudson\altaffilmark{1,2},
Russell J. Smith\altaffilmark{3},
John R. Lucey\altaffilmark{3},
David J. Schlegel\altaffilmark{3,4} \&
Roger L. Davies\altaffilmark{3}
}
\altaffiltext{1}{
  Department of Physics \& Astronomy, University of Victoria,
  P.O. Box 3055, Victoria, B.C. V8W 3P6, Canada.
  E-mail: hudson@uvastro.phys.uvic.ca}
\altaffiltext{2}{Visiting Astronomer, Cerro Tololo Inter-American
  Observatory (CTIO), the National Optical Astronomy Observatories.
  operated by the Association of Universities for Research in
  Astronomy (AURA) under a cooperative agreement with the the National
  Science Foundation (NSF).}
\altaffiltext{3}{Department of Physics, University of Durham, Science
  Laboratories, South Road, Durham DH1 3LE, England. \\ E-mail:
  R.J.Smith, John.Lucey, R.L.Davies@durham.ac.uk}
\altaffiltext{4}{Present address: Dept. of Astrophysical Sciences,
  Peyton Hall, Princeton University,
  USA. E-mail:schlegel@astro.princeton.edu}

\begin{abstract}
We report first results from the ``Streaming Motions of Abell
Clusters'' (SMAC) project, an all-sky Fundamental Plane survey of 699
early-type galaxies in 56 clusters between $\sim{}3000 \kms$ and
$\sim{}14000 \kms$.  For this sample, with a median distance of
$\sim{}8000 \kms$, we find a bulk flow of amplitude $630\pm200\kms$,
towards \lb{260\pm15}{-1\pm12}, with respect to the Cosmic Microwave
Background.  The flow is robust against the effects of individual
clusters and data subsets, the choice of Galactic extinction maps,
Malmquist bias and stellar population effects.  The direction of the
SMAC flow is $\sim$90$^\circ$ away from
the flow found by 
Lauer \& Postman, but
is in good agreement with the gravity dipole predicted from the
distribution of X-ray-luminous clusters.  Our detection of a
high-amplitude coherent flow on such a large scale argues for excess
mass density fluctuation power at wavelengths $\lambda \ga 60$ \hmpc,
relative to the predictions of currently-popular cosmological models.
\end{abstract}


\section{Introduction}
The dipole anisotropy of the Cosmic Microwave Background (CMB)
radiation is generally interpreted as a Doppler effect due to the
motion of the Sun with respect to the CMB rest-frame.  The velocity of
the Local Group (LG), 627$\pm$22\,\kms\ towards \lb{276\pm3}{30\pm3},
is now well determined from COBE (Kogut et al.\ 1993). Less well
known, however, is the depth and degree to which nearby galaxies share
in this motion and, by implication, the scales and amplitudes of the
mass fluctuations responsible for the flow.

To a depth of $\sim{}6000$\,\kms, a rough consensus has emerged from
recent peculiar velocity surveys of galaxies.  For instance,
Giovanelli et al. (1998) find a flow (in the CMB frame) of
$200\pm65$\,\kms\ towards \lb{295}{25}, from an I-band Tully--Fisher
survey, while the Mark III velocity compilation yields
$370\pm110$\,\kms\ towards \lb{306}{13} (Dekel et al.\ 1998, in
prep.).  Beyond this depth, however, the situation is much less clear.
Lauer \& Postman (1994, hereafter LP), using a photometric distance
indicator for brightest cluster galaxies, found a $689\pm178\,\kms$
bulk motion (towards \lb{343}{52}) for an all-sky sample of 119
Abell/ACO clusters to 15000\,\kms\ depth.  Currently-popular
cosmological models have too little large-scale power to generate
coherent flows on such large scales (Feldman \& Watkins 1994; Strauss
et al.\ 1995; Jaffe \& Kaiser 1995). The LP result has been challenged
by subsequent peculiar velocity surveys (Riess et al.\ 1995;
Giovanelli et al.\ 1996; Hudson et al.\ 1997, hereafter H97;
Giovanelli et al.\ 1997; M\"{u}ller et al.\ 1998; Dale et al.\ 1998).
However, with the exception of Dale et al.\ (1998), none of these
studies approach the depth and sky coverage of the LP survey.

In this {\em Letter\/}, we report the detection of a significant
coherent flow from the SMAC survey of cluster peculiar motions.  The
survey comprises 699 early-type galaxies in 56 clusters mainly within
12000\,\kms.  A more detailed description and analysis of the survey
will be presented in a forthcoming series of papers.

\section{Data \& Method}

The distance indicator used in this study is the Fundamental Plane
(FP) of early-type galaxies (Davis \& Djorgovski 1987; Dressler et
al. 1987). The FP relates the effective (half-light) radius, $\log
R_e$ (the distance-dependent quantity), the mean surface brightness
within this radius, $\expec{\mu}_e$, and the central velocity
dispersion, $\log \sigma$.

The SMAC cluster sample consists of new data for 40 clusters (to be
reported in future papers), supplemented with data from the literature
for 16 clusters previously studied in H97.  All but 8 of these are
Abell/ACO clusters, and all but 5 have $cz_{\sun} < 12000 \kms$.  The
median distance of the SMAC sample is $\sim{}8000 \kms$.  The number
of early-type galaxies observed, per cluster, is in the range 4 -- 56,
with a median of 8.

New velocity dispersions and Mg$_2$ linestrengths were obtained from
the Isaac Newton 2.5m and Anglo-Australian 3.9m telescopes.  We
followed the homogenization procedure of Smith et al.\ (1997) to bring
new spectroscopic data and existing data from the literature onto a
common system.  Corrections of up to 0.03 dex in $\sigma$ are derived
for each system, but typical uncertainties in these corrections are
only 0.005--0.008 dex, corresponding to 1.6--2.6\% systematic
uncertainty in distance, per observing run.  Of the 699 galaxies in
the final SMAC sample, velocity dispersion data is drawn wholly from
new observations for 41\% of the sample. For 21\% of the sample, the
final $\sigma$ derives from both new data and published
measurements. For the remaining 38\%, the dispersions are derived from
previously published data.

New R-band photometric data were obtained from the Jacobus Kapetyn
1.0m and Cerro Tololo Inter-American Observatory 0.9m telescopes.
Effective radii and surface brightnesses were determined by fitting a
one-dimensional $R^{1/4}$-law profile to the circular aperture
photometry, with corrections for seeing, cosmological effects and
Galactic extinction (using the maps of Schlegel, Finkbeiner and Davis
1998; hereafter SFD).  Additional photometric data has been drawn from
the literature and carefully combined with newly obtained parameters.
Our new photometry provides data for roughly half of the 699 galaxies
in the final SMAC sample.

Our method for determining cluster distances and bulk flows follows
H97.  We summarize the important points here.  To obtain cluster
distances, we use the inverse form of the FP, i.e.\ we regress on the
distance-independent quantity $\log \sigma$. This regression is
insensitive to photometric selection effects (e.g.\ selection on
magnitude or diameter; see H97, Strauss \& Willick 1995).  The FP
slopes and scatter for the SMAC sample are consistent with previous
results for the inverse FP (H97). The resulting distance error is 21\%
per galaxy, and 3\% -- 11\% per cluster, with a median of 7\%.  We
make a correction for Malmquist bias under the assumption that
clusters are drawn from a homogeneous underlying density field.  Due
to the large number of objects per cluster, the Malmquist bias
corrections are small (typically $\sim{}2\%$) for this sample.
Finally, we adjust the FP zero-point so that the cluster peculiar
velocities have no net radial inflow or outflow with respect to the
CMB frame.

A bulk flow model provides the simplest parameterization of the
peculiar velocity field. We minimize
\begin{equation}
\chi^2 = \sum_i \frac{(v_i - {\bf V} \cdot \hat{\bf r}_i)^2}{\epsilon^2_i}
\label{eq:chi}
\end{equation}
where $v_i$ is the CMB frame peculiar velocity of cluster $i$ with
direction vector $\hat{\bf r}_i$, and ${\bf V}$ is the bulk flow. The
total error, $\epsilon_i$, is the quadrature sum of the cluster
distance error (in the range 100--1200\,\kms), the error in the mean
cluster $cz$ (typically 150\,\kms) and a ``thermal component'' (set to
250 \kms) which allows for small scale fluctuations around the mean
bulk flow. Given the error-weighting in \eqref{chi}, the weighted mean
depth of the sample is $\sim{}6700 \kms$.

We have tested the bulk flow recovery with Monte Carlo simulations in
which galaxies are assigned peculiar velocities consistent with an
input bulk flow plus random errors.  Because our survey has good
sky-coverage, we find that our bulk flow fit is not affected by
``geometry bias'', i.e.\ there is little covariance between the
monopole and dipole (bulk flow) terms.

\section{Results}

The principal result of this {\em Letter\/} is that the SMAC sample
exhibits a CMB-frame bulk flow of $V = 630\kms$ (error-bias corrected,
see LP), towards \lb{260\pm15}{-1\pm12}.  The peculiar velocities of
the sample clusters are shown, as a function of angle from the apex
direction, in \figref{cos}.  The linear trend seen here is a clear
signature of a bulk streaming motion.

The random error in the bulk flow due to distance and velocity errors
is 180 \kms.  A further uncertainty in the bulk flow arises from
uncertainties in the corrections to $\sigma$ for different runs.  We
estimate this uncertainty by generating bootstrap realizations of the
corrections and analyzing the bootstrap-corrected datasets in the same
way as the real data.  Through this procedure, we find that this
source of systematic error contributes an uncertainty of 90 \kms\ to
the measured bulk flow.

The total velocity error in the direction of the flow is 200 \kms.
The error ellipsoid is triaxial: the direction of smallest error
($\pm100 \kms$) is along \lb{314}{49}, and the direction of largest
error ($\pm 214 \kms$) is along \lb{237}{-11}, which is within
24\arcdeg\ of our bulk flow.  Allowing for the three degrees of
freedom in the bulk flow vector we find that our sample is
inconsistent with being at rest in the CMB frame at the 99.9$\%$
confidence level.  A 95$\%$ lower limit on the bulk motion is 400
\kms.  If we divide the sample by distance into two parts with equal
errors, the outer shell of the SMAC sample gives a larger amplitude
flow (by 240$\pm$360\,\kms) but is consistent, within the errors, with
that found from the inner sphere.  Both inner and outer samples
independently yield a bulk flow with a significance $\ga 98\%$.

\vbox{%
\begin{center}
\leavevmode
\hbox{%
\epsfxsize=8.9cm
\epsffile{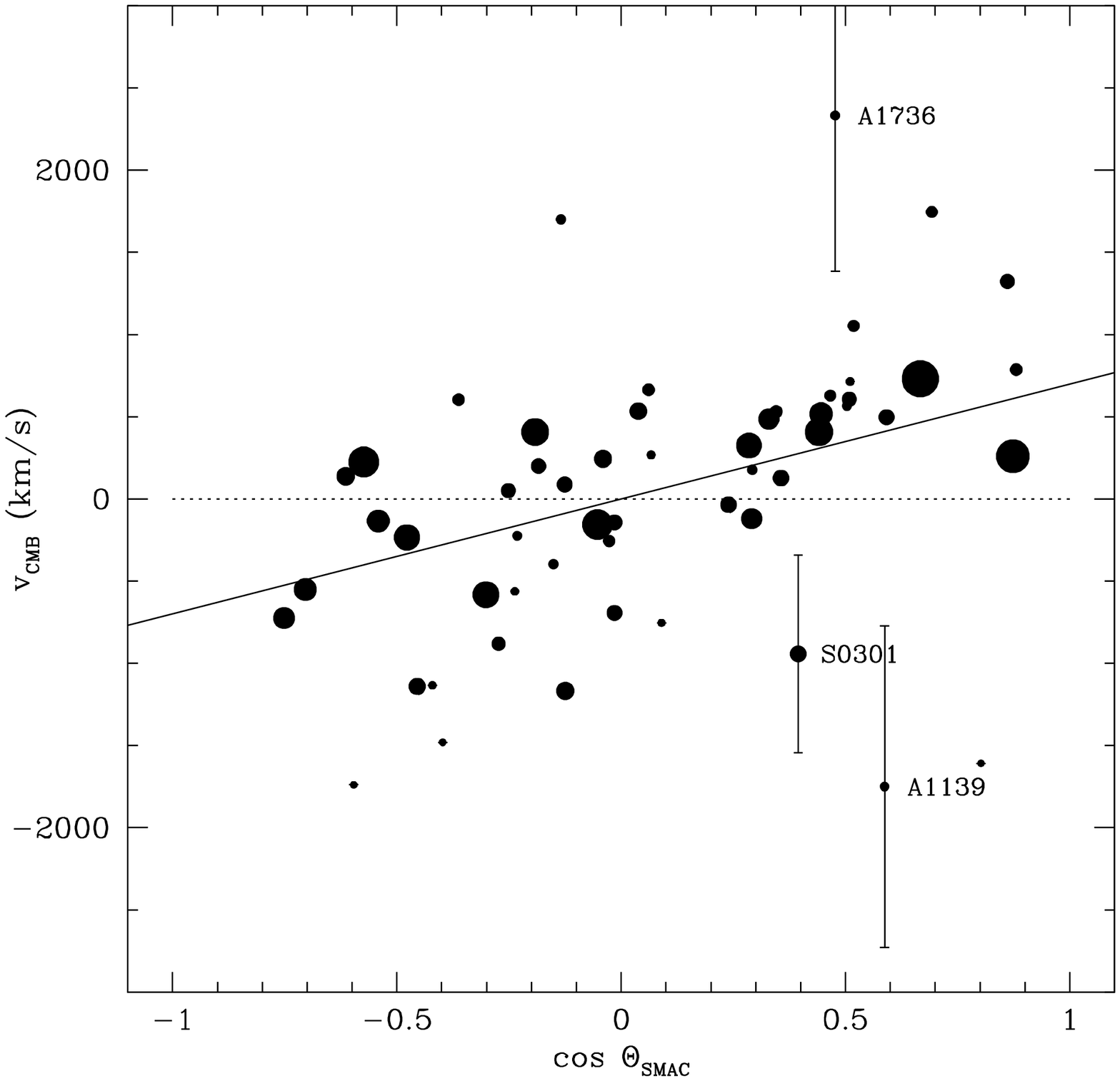}}
\begin{small}
\figcaption{%
Peculiar velocities as a function of angle from the apex of the bulk
flow. Symbol sizes are inversely proportional to errors.  The solid
line is the best-fitting bulk flow model, and clusters which deviate
from this flow at the $2\sigma$ level are indicated, with error bars.
\label{fig:cos}}
\end{small}
\end{center}}

We have explored a wide range of potential systematic effects.  By
excluding each cluster from the sample in turn, we find that no
individual cluster is responsible for more than 50 \kms\ of the total
bulk motion.  No single supercluster structure dominates the flow.  We
find a similar result excluding individual spectroscopic runs.
Furthermore, because of the balanced sky coverage, the bulk flow is
insensitive to the adopted zero-point of the FP relation.  A 1\%
change in the zero-point (corresponding to its $1\sigma$ uncertainty)
alters the bulk flow by only $\sim{}25 \kms$.  If, instead of SFD
extinction corrections, we use those of Burstein \& Heiles (1982), the
recovered bulk flow drops by $\sim$ 100\,\kms.  In order to assess the
effect of Malmquist bias and redshift cuts, we have performed a
simultaneous FP and bulk flow fit (``Method II'' in the terminology of
Strauss \& Willick 1995); we recover a bulk flow which is identical
within the errors ($\sim$ 40\,\kms\ change).  The effect of a possible
FP dependence on morphological type is negligible: E and S0 subsamples
of the data give consistent results (they differ by
$140\pm330$\,\kms).  Finally, we have examined the possible effect of
cluster-to-cluster stellar population differences, using the \mg\
index as an age/metallicity indicator.  We find that our clusters are
consistent with following a universal \mgsig\ relation. Furthermore,
we find no correlation between peculiar velocity and the
(non-significant) offsets of clusters from the \mgsig\ relation.

\vbox{%
\begin{center}
\leavevmode
\hbox{%
\epsfxsize=8.9cm
\epsffile{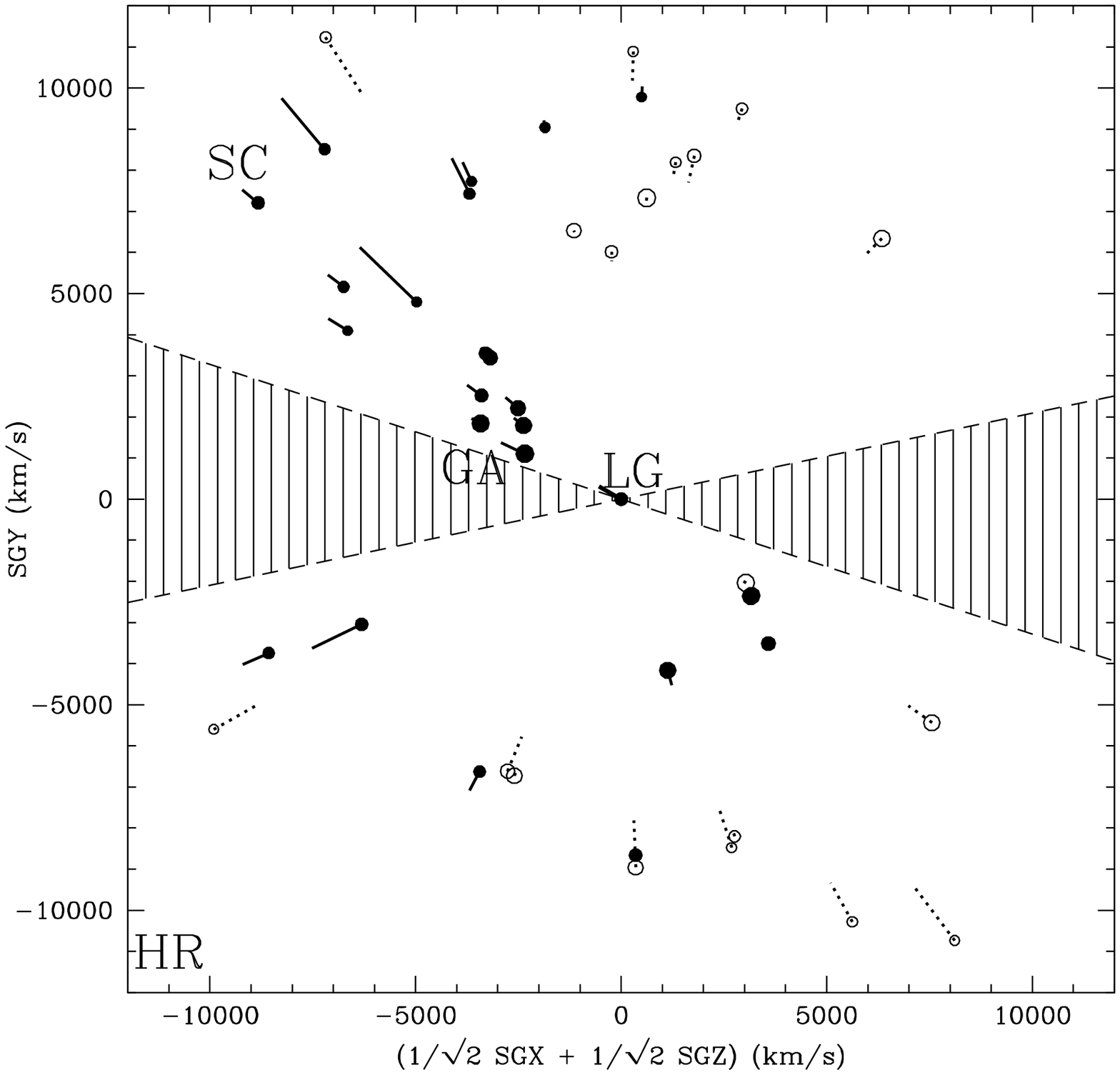}}
\begin{small}
\figcaption{%
Peculiar velocities projected onto a plane between SGX=0 and
SGZ=0. The negative half of the horizontal axis points towards
\lb{272}{-3}, which is only $\sim{}13\arcdeg$ from the direction of
the SMAC bulk flow.  We plot only the 43 clusters within 45$^\circ$ of
the plane. The shaded region indicates the Zone of Avoidance ($|b| <
15$). The distance to each cluster is indicated by the circle and the
CMB frame redshift is indicated by the tip of the vector. Outflowing
clusters have filled circles and solid lines; inflowing ones have open
circles and dotted lines. Circle size is inversely proportional to the
error in the 
peculiar velocity. 
Important locations are indicated: the Local
Group (LG); the Great Attractor (GA); the Shapley Concentration (SC)
and the Horologium-Reticulum (HR) supercluster.
\label{fig:plane}}
\end{small}
\end{center}}

\section{Discussion}

The motion of the Local Group (LG) and the bulk motion of the SMAC
sample both lie in the plane which is at a 45\arcdeg angle from the
SGX=0 and SGZ=0 Supergalactic planes. The CMB-frame peculiar
velocities of SMAC clusters within $\pm45\arcdeg$ of this plane are
shown in \figref{plane}.  This figure shows the continued outflow
beyond the Great Attractor (``GA'', Lynden-Bell et al.\ 1988) both
above and below the Galactic Plane, indicating that the local motions
are not generated wholly by the GA.  On the opposite side of the sky
the data, though sparser, exhibit an inflow of similar magnitude,
suggesting a bulk flow with very large coherence length.

\begin{figure*}
\figurenum{3}
\plotone{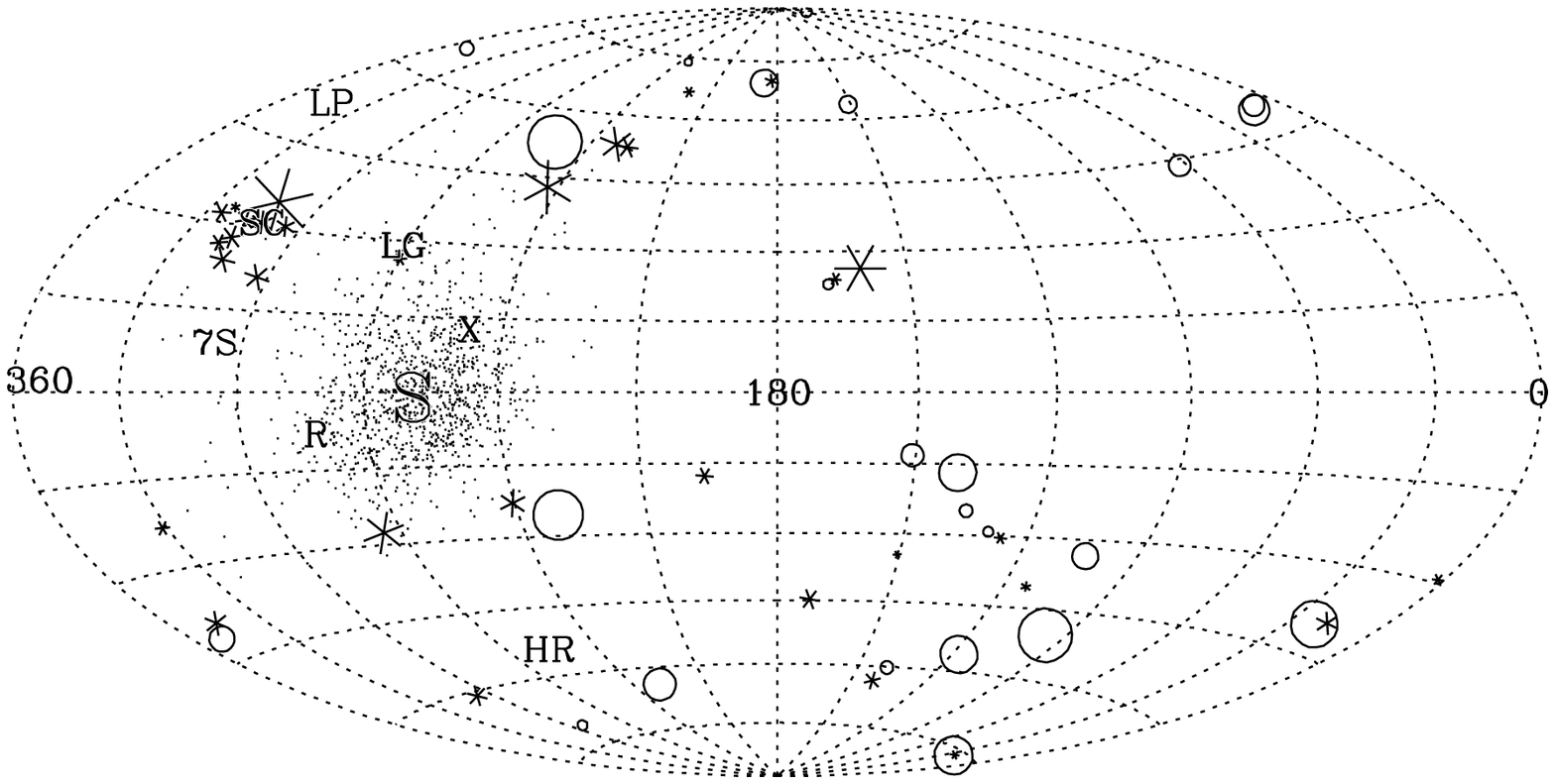}
\caption{%
Peculiar velocities and bulk flow directions in Galactic coordinates
(Aitoff projection). Clusters in the SMAC survey are indicated by open
circles (inflowing) or asterisks (outflowing), with symbol size
proportional to peculiar velocity. The direction of the SMAC bulk flow
is indicated by the large ``S'', while the dots show 1000 directions
drawn from the bulk flow error ellipsoid.  Important directions are
indicated: the motion of the Local Group (LG); the motion of the
Lynden-Bell et al. (1988) sample (7S); the bulk flow of LP; the motion
of the Riess et al.\ (1997) SNIa sample (R); the Shapley Concentration
(SC); the Horologium-Reticulum supercluster (HR); and the predicted
motion of the LG from X-ray clusters (X) (Plionis \& Kolokotronis
1998).  The bulk flow of Dale et al.\ (1998) is small and consistent
with zero.  Consequently, its direction is not meaningful and has not
been plotted here.
\label{fig:aitoff}
}
\end{figure*}

The direction of the SMAC bulk flow in Galactic coordinates is shown
in \figref{aitoff}.  Also shown are the positions of two most
prominent concentrations of clusters within 300 \hmpc\ (Tully et al.\
1992), namely the Shapley Concentration (Scaramella et al.\ 1989;
Raychaudhury 1989) and the Horologium-Reticulum supercluster (Lucey et
al.\ 1983). The direction of the SMAC bulk motion is roughly between
these two concentrations.  Indeed, there is some evidence in our
sample of inflow towards the Shapley Concentration.  The data are too
sparse in the foreground of Horologium-Reticulum to measure any
infall.  The direction of our flow agrees very well ($< 16$\arcdeg)
with the direction of LG motion predicted from the density field of
X-ray/Abell clusters (Plionis \& Kolokotronis 1998).  
The IRAS PSCz galaxy redshift survey is of sufficient depth to yield  
accurate predicted peculiar velocities (Branchini et al.\ 1998) for   
our clusters.  A preliminary comparison indicates excellent           
directional agreement between the PSCz-predicted and the observed bulk 
flow of SMAC clusters.                                                

How does our bulk flow compare to results from other surveys?  One
might naively estimate the level of agreement between two surveys from
a $\chi^2$ statistic derived from the difference between the two bulk
flow 
vectors and the sum of their observational-error covariance matrices.
This procedure is incorrect, however, because it does not allow for
the fact that different surveys probe different volumes and regions on
the sky.  As stressed by Watkins \& Feldman (1995), the incomplete
cancelation of small-scale flows internal to a volume can have a
significant effect on the total measured bulk flow.  The significance
of disagreements between flow vectors is always {\em overestimated} if
these effects are neglected.

For example, according to the naive comparison described above, our
result is in conflict with the bulk flow 
vector 
of LP and of Dale et al.\
(1998).  Our flow is of similar amplitude to that found by LP, but the
direction is $\sim{}90\arcdeg$ from the apex of the LP dipole, so at
face value the agreement is poor.  Dale et al.\ find a bulk flow of
$\la 200 \kms$ for their deep Tully--Fisher cluster sample, in
contrast to our result%
\footnote{It is worth noting, however, that our error-weighted bulk
flow is within the $2\sigma$ error ellipse of Dale et al.'s
volume-weighted solution, although not of their error-weighted
solution.}.  A detailed analysis, allowing for the effects of survey
geometries and internal flows, is required in order to determine
whether the apparent disagreements with LP and Dale et al.\ are
in fact significant.  Such an analysis is in progress.

In contrast to the two surveys just discussed, our flow agrees very
well in amplitude and direction with that of the similarly deep (9000
-- 13000 \kms) Tully--Fisher sample of Willick (1998, also private
communication).  The SMAC bulk flow is not inconsistent with
preliminary results from the EFAR project (Saglia et al. 1998).
Finally, we have performed a bulk flow fit to the SNIa velocity data
of Riess et al.\ (1997), which yields a flow of $320\pm160$\,\kms\
(error-bias corrected) directed towards \lb{282}{-8}.  This direction
is only $\sim{}20^\circ$ from the SMAC flow apex.

Our result can be compared to expectations based on popular families
of cosmological models.  The bulk flow amplitude measured by a typical
(i.e. randomly-placed) observer depends on the power spectrum of mass
density fluctuations and the geometry of the survey.  Following Kaiser
(1988), we have calculated the $k$-space window function for the bulk
flow given the SMAC survey geometry.  We find that the SMAC bulk flow
probes the mass power spectrum on scales larger than $\lambda \sim{}60
\hmpc$.  If we fix the power spectrum on the largest scales using
COBE, then the high amplitude of the SMAC bulk flow implies
substantial power on intermediate scales.  For variants of the cold
dark matter (CDM) cosmology with either a cosmological constant,
neutrinos, or a tilt ($n \neq 1$) in the initial primordial spectrum,
the choice of free parameters fixes the power spectrum on 
sub-COBE 
scales.  For the three COBE-normalized CDM variants, we require
choices of the parameters such that the rms fluctuation in mass
density contrast in an 8 \hmpc\ sphere is $\sigma_8\,\Omega^{0.6} \ga
0.87$, 0.80 or 0.85, respectively, at the 90\% confidence level.  This
is in conflict with the value $\sigma_8\Omega^{0.5} \sim{}0.55$
inferred from the abundance of rich clusters (Vianna \& Liddle 1996;
Eke et al.\ 1996; Pen 1998), but is in better agreement with the value
$\sigma_8 \Omega^{0.6} \sim{}0.8$ inferred from other peculiar
velocity surveys (Kolatt \& Dekel 1997; Zaroubi et al.\ 1997).
Alternatively, if the power spectrum is to fit {\em both\/} the
abundance of rich clusters {\em and\/} the COBE fluctuations, then it
must be considerably more ``peaked'' between these scales, relative to
the CDM variants.

\section{Summary}

We have recently completed a FP survey of 699 early-type galaxies in
56 clusters within $\sim{}12000 \kms$.  For this sample, we find a
large-scale bulk flow of amplitude $630\pm200\kms$ towards
\lb{260\pm15}{-1\pm12}.  Our result is robust against the effects of
individual clusters and data subsets, the choice of Galactic
extinction maps, Malmquist bias and stellar population effects.

This result suggests that the mass fluctuation power spectrum has a
high amplitude on scales $\lambda \sim{}60 - 600 \hmpc$.  This regime,
near the peak of the power spectrum, is at present poorly constrained
by other methods.  On these scales, CMB anisotropies are dominated by
acoustic oscillations and depend mainly on the baryon content, whereas
current galaxy redshift surveys are insufficiently deep, and are
subject to the (unknown) relationship of galaxies to the underlying
mass density fluctuations.

Further analyses, currently underway, will (i) determine the
consistency of our results with those of other surveys, allowing for
the different sample geometries; (ii) compare measured velocities with
predictions of the IRAS PSCz redshift survey; and (iii) constrain
directly the mass power spectrum, by means of the dipole and higher
moments of the velocity field.

\acknowledgments 
We thank the staff of the Isaac Newton Group of
telescopes at the Observatorio del Roque de los Muchachos, the
Anglo-Australian Observatory and the Cerro Tololo Inter-American
Observatory, where these observations were conducted.

\end{document}